\documentclass{article}

\usepackage{arxiv}

\usepackage{amssymb}
\usepackage{amsmath}
\usepackage{lineno}
\usepackage{textcomp}
\usepackage{float}  
\usepackage{caption}
\usepackage{subcaption}
\usepackage{indentfirst}
\usepackage{enumerate}
\usepackage{microtype}
\usepackage{chemformula}
\usepackage{setspace}
\usepackage{yfonts,color}
\usepackage{siunitx}
\usepackage{comment}
\usepackage{multirow}
\usepackage{varioref}
\usepackage[bottom]{footmisc}
\usepackage{wrapfig}
\usepackage{float}
\usepackage{type1cm}
\usepackage{chngcntr}
\usepackage[nottoc, notlof, notlot]{tocbibind}
\usepackage{hyperref}
\usepackage{subfiles}
\usepackage{soul} 
\usepackage{graphicx}

\begin{document}


\title{Time-based Selection of Kaonic Atom X-ray Events with Quasi-Hemispherical CZT Detectors at the DA$\Phi$NE collider}

\maketitle

Francesco Artibani$^{1,2,*}$,
Leonardo Abbene$^{3,1}$,
Antonino Buttacavoli$^{3,1}$,
Manuele Bettelli$^4$,
Gaetano Gerardi$^{3}$,
Fabio Principato$^{3,1}$,
Andrea Zappettini$^4$,
Massimiliano Bazzi$^1$,
Giacomo Borghi$^{5,6}$,
Damir Bosnar$^{7,1}$,
Mario Bragadireanu$^8$,
Marco Carminati$^{5,6}$,
Alberto Clozza$^1$,
Francesco Clozza$^{1,10}$,
Raffaele Del Grande$^{11,1}$,
Luca De Paolis$^1$,
Carlo Fiorini$^{5,6}$,
Ivica Friscic$^7$,
Carlo Guaraldo$^1$\textsuperscript{\textdagger},
Mihail Iliescu$^1$,
Masahiko Iwasaki$^{14}$,
Aleksander Khreptak$^{12,13,1}$,
Simone Manti$^1$,
Johann Marton$^9$,
Pawel Moskal$^{12,13}$,
Fabrizio Napolitano$^{15,16,1}$,
Hiroaki Ohnishi$^{17}$,
Kristian Piscicchia$^{18,1}$,
Francesco Sgaramella$^1$,
Michal Silarski$^{12}$,
Diana Laura Sirghi$^{1,8,18}$,
Florin Sirghi$^{1,8}$,
Magdalena Skurzok$^{12,13,1}$,
Antonio Spallone$^1$,
Kairo Toho$^{17,1}$,
Lorenzo G. Toscano$^{5,6}$,
Oton Vazquez Doce$^1$,
Johann Zmeskal$^9$\textsuperscript{\textdagger},
Catalina Curceanu$^1$,
Alessandro Scordo$^1$

{$^1$Laboratori Nazionali di Frascati, INFN, Via E. Fermi 54, 00044 Frascati, Italy} \\
{$^2$Dipartimento di Matematica e Fisica, Università di Roma Tre, Via della Vasca Navale 84, 00146 Roma, Italy} \\
{$^3$Dipartimento di Fisica e Chimica - Emilio Segrè, Università di Palermo, Viale Delle Scienze Edificio 18, 90128 Palermo, Italy} \\
{$^4$Istituto Materiali per l’Elettronica e il Magnetismo, Consiglio Nazionale delle Ricerche, Parco Area delle Scienze 37/A, 43124 Parma, Italy}\\
{$^5$Dipartimento di Elettronica, Informazione e Bioingegneria, Politecnico di Milano, Piazza Leonardo da Vinci 32, 20133 Milano, Italy}\\
{$^6$INFN Sezione di Milano, Via Giovanni Celoria 16, 20133 Milano, Italy}\\
{$^7$Department of Physics, Faculty of Science, University of Zagreb, Bijenička cesta 32, 10000 Zagreb, Croatia}\\
{$^8$Horia Hulubei National Institute of Physics and Nuclear Engineering (IFIN-HH), No. 30, Reactorului Street, 077125 Magurele, Ilfov, Romania}\\
{$^9$Stefan-Meyer-Institut für Subatomare Physik, Dominikanerbastei 16, 1010 Wien, Austria}\\
{$^{10}$Dipartimento di Scienze Matematiche Fisiche e Naturali, Università degli Studi di Roma Tor Vergata, Via Cracovia 90, 00133 Roma, Italy}\\
{$^{11}$Faculty of Nuclear Sciences and Physical Engineering, Czech Technical University in Prague, Břehová 7, 115 19 Prague, Czech Republic}\\
{$^{12}$Faculty of Physics, Astronomy, and Applied Computer Science, Jagiellonian University, Lojasiewicza 11, 30-348 Krakow, Poland}\\
{$^{13}$Centre for Theranostics, Jagiellonian University, Kopernika 40, 31-501 Krakow, Poland}\\
{$^{14}$RIKEN, 2-1 Hirosawa, Wako, Saitama 351-0198, Japan}\\
{$^{15}$Via A. Pascoli 06123, Perugia (PG), Italy, Dipartimento di Fisica e Geologia, Università degli studi di Perugia}\\
{$^{16}$INFN Sezione di Perugia, Via A. Pascoli – 06123 Perugia – Italia}\\
{$^{17}$Research Center for Accelerator and Radioisotope Science (RARIS), Tohoku University, 1-2-1 Mikamine, Taihaku-ku, 982-0826 Sendai, Japan}\\
{$^{18}$Centro Ricerche Enrico Fermi - Museo Storico della Fisica e Centro Studi e Ricerche “Enrico Fermi”, Via Panisperna 89A, 00184 Roma, Italy}\\
{\textsuperscript{\textdagger} Deceased.}\\

{$^*$Author to whom any correspondence should be addressed.}

{francesco.artibani@infn.it, francesco.artibani@uniroma3.it}



\begin{abstract}

{\noindent This work presents the results of a time-based event selection for searching X-ray signals from kaonic atom X-ray transition using a single quasi-hemispherical Cadmium-Zinc-Telluride (CZT) detector at the DA$\Phi$NE collider. To mitigate the high background level in the measured X-ray spectrum, a dedicated event selection strategy was developed, exploiting the precise timing correlation between e$^+$e$^-$ collisions and detector signals. 

\noindent This approach enabled, for the first time, the observation of two characteristic X-ray transitions from kaonic aluminum atoms using a CZT detector: for the 5–4 transition at 50~keV, 362~$\pm$~41~(stat.)~$\pm$~20~(sys.) signal events over 1698~$\pm$~197~(stat.)~$\pm$~25~(sys.) background events in 5$\sigma$ were observed, with a resolution of 9.2\%~FWHM; for the 4–3 transition at 106~keV, 295~$\pm$~50~(stat.)~$\pm$~20~(sys.) signal events over 2939~$\pm$~500~(stat.)~$\pm$~16~(sys.) background events in 5$\sigma$ were measured, with a resolution of 6.6\%~FWHM. A strong background suppression of approximately 95\% of the triggered data was achieved through this time-based selection.

\noindent The demonstrated timing capability of the CZT detector proved highly effective in isolating time-correlated events within an 80~ns window, setting an important benchmark for the application of these semiconductors in timing-based X-ray spectroscopy.

\noindent These results highlight the potential of CZT-based detection systems for future precision measurements in high-radiation environments, paving the way for compact, room-temperature X-ray and $\gamma$-ray spectrometers in fundamental physics and beyond.}

\end{abstract}

\section{Introduction}

\noindent Nuclear and particle physics experiments require detectors capable of covering a broad energy range while maintaining high energy resolution. Additionally, good timing performance, environmental stability, and resilience to high radiation fluxes over extended periods, especially in collider-based experiments, are essential characteristics for such applications.

\noindent In recent years, significant efforts have been devoted to developing compound semiconductors with wide band gaps and high atomic numbers for X and $\gamma$-ray detection. One of the main advantages of these materials lies in the possibility of tailoring their physical properties, such as band gap, atomic number, and density, making them highly adaptable to a broad range of applications.

\noindent Among compound semiconductors for radiation detection, Cadmium Zinc Telluride (CZT, CdZnTe) has been confirmed as one of the most promising \cite{del_sordo_progress_2009}. This material exhibits high detection efficiency, good energy and time resolutions at room temperature, up to hundreds of keV, making it ideal for experiments in both applied and fundamental research using compact, non-invasive setups \cite{iniewski_czt_2014, tang_cadmium_2021}. Recent progress in crystal growth, as well as in innovative techniques for electric contacts and data processing \cite{Abbene_energyres_2013,abbene_development_2017,abbene_room-temperature_2020,Vicini:2023opt}, has made possible to develop new precise systems for X-ray and $\gamma$-ray detection using this technology.

\noindent The first application of a CZT-based detection system in fundamental particle and nuclear physics was carried out by the SIDDHARTA-2 (Silicon Drift Detectors for Hadronic Atom Research by Timing Application) collaboration \cite{Curceanu:2019uph, Artibani:2024kop, Curceanu:2023yuy} at the DA$\Phi$NE collider at Frascati National Laboratories of Istituto Nazionale di Fisica Nucleare (INFN-LNF) by testing, characterizing and using for the first time a CZT-based apparatus to perform intermediate-mass kaonic atoms spectroscopy \cite{Abbene:2023vhm, Scordo:2023per, artibani2024NewOpp, Artibani:2025tem}.

\noindent The experiment exploits the DA$\Phi$NE $\phi$-factory \cite{Milardi:2009zza,Milardi:2018sih, Milardi:2021khj, Milardi:2024efr} to produce a low-energy charged kaon beam, originating from the decay of the $\phi$ meson. The kaons are then stopped in a target so that they can bind to a nucleus and form an atomic bound state. The kaonic atom is formed in a highly excited state and undergoes a cascade de-excitation, which is radiative in its final stages. By performing spectroscopic measurements of these transitions, relevant observables can be extracted, crucial for a wide range of topics in fundamental physics, from low-energy strong interaction studies \cite{Curceanu:2019uph, Curceanu:2020kkg, Curceanu:2023yuy, Cieply:2016jby,Obertova:2022des} and astrophysics \cite{Merafina:2020ffb, Tolos:2020aln}, to atomic cascade model validations \cite{Hartmann1990, Gotta:2004rq} and precision tests of QED \cite{Sgaramella:2024klx}.


\noindent This work represents a significant step forward in demonstrating the feasibility of using this technology under the harsh conditions common in fundamental physics experiments, such as the high radiation flux to which the detectors are exposed and the need to extract signals several orders of magnitude smaller than the background. The deployment of CZT detectors in such settings opens new prospects for compact, room-temperature, and high-resolution detection systems in fundamental physics research.

\noindent One of the primary challenges in this application is the suppression of intense background radiation. This elevated background level, typical of collider environments characterized by extremely high event rates, arises from two main sources: beam losses occurring within the accelerator beam pipes (asynchronous background), and radiation directly produced by particle collisions at the interaction point (synchronous background).

\noindent Mitigating the impact of this background requires the development of a detection system capable of discriminating the signal from both synchronous and asynchronous background sources. This necessitates not only dedicated hardware, but also a data selection strategy that fully exploits the timing capabilities of the CZT detectors of the collaboration, allowing for a precise temporal separation between signal and background events. Furthermore, the integration of complementary detection systems operating in coincidence with the CZT modules is essential. These additional systems provide an extra layer of time-based selection, further enhancing the robustness and reliability of the measurement strategy.

\noindent In this paper, we present the experimental setup for the  CZT-based detectors and the key aspects of the data selection performed on the first dataset collected at DA$\Phi$NE during the 2024 data-taking campaign. Specifically, the data discussed here were acquired using a single CZT detector between February 28$^{\text{th}}$ and March 20$^{\text{th}}$, 2024, with a total delivered integrated luminosity of {81~pb$^{-1}$}, using an aluminium target.

\noindent Section 2 provides an overview of the experimental setup. Section 3 outlines the key steps of the data selection procedure applied to one of the detectors, including relevant findings that confirm the fast timing performance achieved by the CZT detectors. Finally, in Section 4, we present a representative energy spectrum from a single run, demonstrating both the effectiveness of the analysis strategy and, for the first time, the observation of kaonic atom transitions with a CZT detector.

\section{Experimental Setup}

\subsection{DA$\Phi$NE Collider} \label{sec: dafne coll}

\begin{figure} [H]
    \centering
    \includegraphics[width=0.75\linewidth]{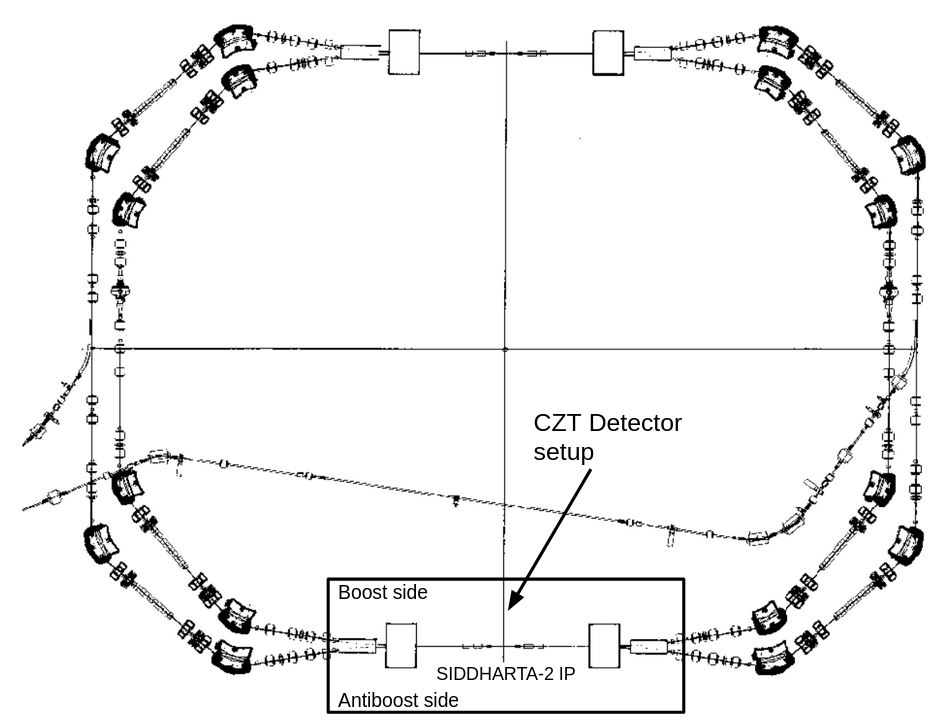}
    \caption{Schematic areal view of the main rings of DA$\Phi$NE collider. The SIDDHARTA-2 interaction point, together with the CZT detection system setup, is highlighted. Adapted from \cite{Bassetti:DAFNERING}.}
    \label{fig: DAFNE scheme}
\end{figure}

\noindent The test measurements were performed at the DA$\Phi$NE collider at INFN-LNF. DA$\Phi$NE is a unique double-ring electron-positron collider operating at a centre-of-mass energy of 1.020 GeV. At this energy, the cross-section for e$^+$e$^-$ inelastic scattering is dominated by $\phi$ meson production. The $\phi$ meson is produced with a slight boost along the axis, directed from the interaction point (IP) inward toward the centre of the rings (boost-side), and promptly decays into a pair of back-to-back low-energy charged kaons ($\beta \approx 0.25$), with a branching ratio (BR) of 48\%. The low-energy negatively charged kaons can thus be easily stopped in a target material, enabling the study of kaonic atoms through the measurement of their cascade transitions.
\noindent A schematic overview of the main rings of the DA$\Phi$NE collider, indicating the SIDDHARTA-2 IP and the location of the CZT detection system used for the measurements described in the following section, is shown in Figure \ref{fig: DAFNE scheme}.

\noindent In the DA$\Phi$NE collider, the dominant source for the machine-induced experimental background are the effects from intra-bunch electromagnetic interactions producing off-energy particles, which arise when high-density beams are operated at relatively low energies to maximize luminosity (Touschek effect) \cite{Boscolo:2007zzb}. A significant fraction of these particles exhibit momentum deviations large enough to prevent them from remaining confined within the nominal beam trajectory. Consequently, they are lost in the focusing magnets before reaching the interaction point, exiting the beam line, and generating a strong electromagnetic background. The CZT detection system is located near the focusing magnets in front of the IP, and in this region, the background is particularly intense, posing a major challenge for precision measurements and requiring an accurate analysis for background reduction.

\subsection{The CZT Detection System}

\noindent The SIDDHARTA-2 experiment was installed at the only active interaction point of the DA$\Phi$NE collider, highlighted in Figure \ref{fig: DAFNE scheme}. The main experimental setup employed Silicon Drift Detectors (SDDs) \cite{Miliucci:2021wbj, Miliucci:2022lvn, Khreptak2023EfficiencyAA} to study kaonic atom transitions from light-mass, cryogenic targets, with X-ray energies ranging from a few keV up to several tens of keV. A detailed description of the setup can be found in \cite{Sirghi:2023wok}.

\noindent The primary SIDDHARTA-2 apparatus was arranged above and below the IP, featuring a sophisticated system composed of multiple detectors. Due to the limited space available only in the plane of the collider’s beam pipe, a compact detection system was required to extend the measurements to kaonic atom transitions of intermediate-mass elements, which emit X-rays in the energy range from tens to hundreds of keV. For this purpose, CZT-based detectors were employed to test and demonstrate the effectiveness of the new technology in this field and to complement the SDD measurements.

\begin{figure} [H]
    \centering
    \includegraphics[width=0.75\linewidth]{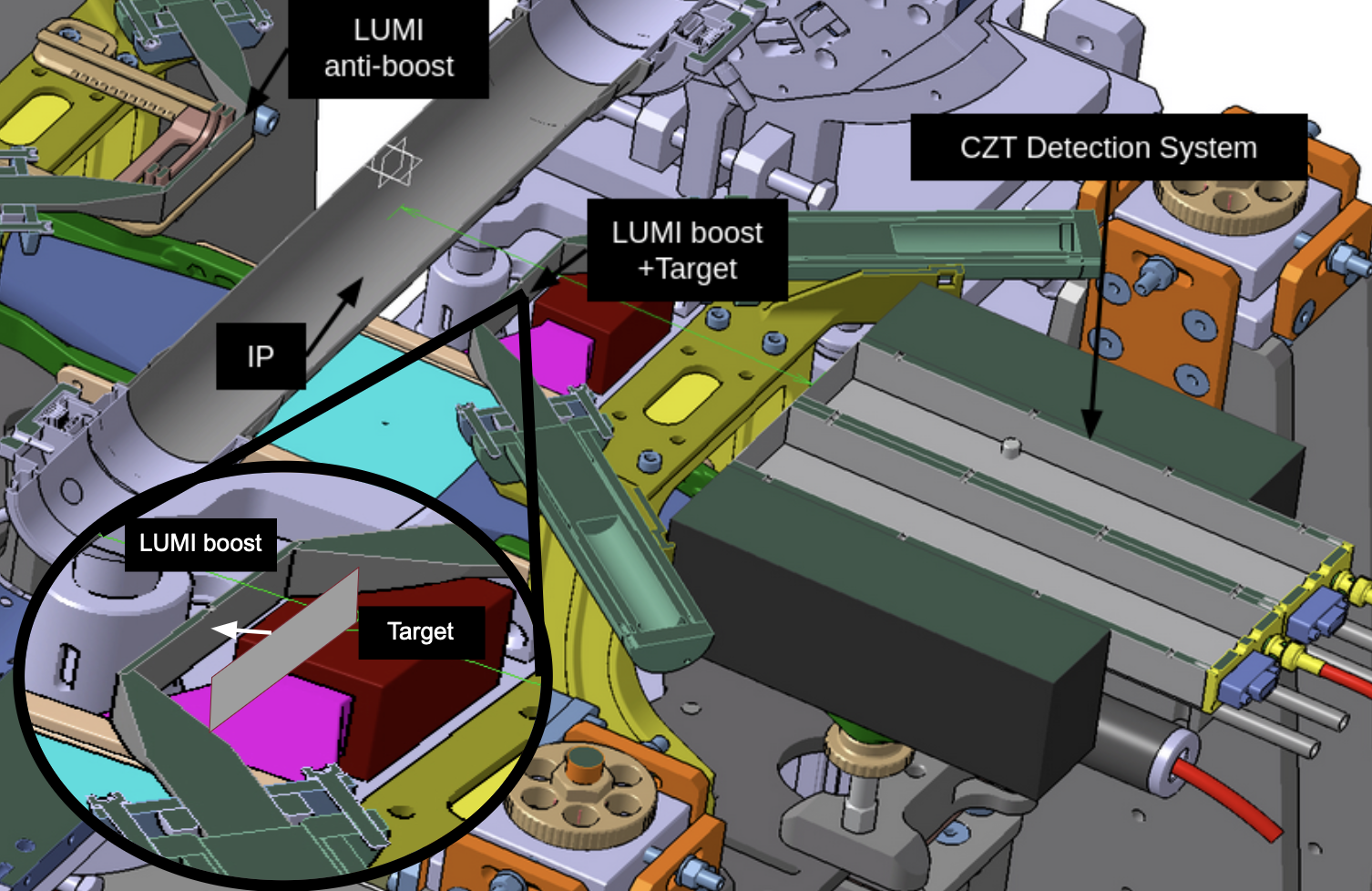}
    \caption{Schematic view of the experimental setup of the CZT detection system in DA$\Phi$NE. A description of the materials follows: The DA$\Phi$NE beam pipe (around the IP in the Figure, grey) is made by a mixture of aluminum and carbon fiber; the LUMI detectors (housing and lightguides shown wrapped in aluminized mylar shown in silver); the Target (shown in light gray) is an aluminum plate; the CZT detection system’s box (shown in light grey) is made of aluminum; mechanical supports and mounting structures (brackets yellow, kinematic mounts orange) are made of aluminum alloy. Additional small components visible: an electronics/power module (magenta) and support plates (cyan/turquoise).}
    \label{fig: CZT setup}
\end{figure}

\noindent During the data-taking campaign, the CZT detectors were placed in the boost side of DA$\Phi$NE's IP, on the plane of the collider rings. Moving outward from the interaction point, the boost-side luminometer (LUMI) was positioned first. This module consists of an 80 mm × 40 mm × 2 mm Scionix BC-408 organic scintillator coupled to two PMTs \cite{Skurzok:2020phi} through plastic lightguides. Used in combination with an identical module on the antiboost side, the LUMI served both for luminosity measurements and for charged kaon selection via time-of-flight (see Section \ref{sec: data analysis}), and was located 10.2 cm from the IP. Immediately after the LUMI, a solid target was placed, in which the kaons stopped and formed atomic states with the nuclei of the target material. In the run presented in this work, an aluminum target was used. Finally, the X-rays emitted by the kaonic atoms were detected by the CZT system, {composed of eight 13 mm × 15 mm × 5 mm quasi-hemispherical CZT detectors grown using Traveling Heater Method (THM) by REDLEN Technology}, enclosed in a thin aluminum box with a 0.27 $\mu$m thick aluminum window, housing both the active detector material and the front-end electronics. The detectors were positioned 17 cm from the target. A schematic view of the setup is shown in Figure \ref{fig: CZT setup}.

\noindent Each CZT detector in the system was connected to a custom front-end electronic chain based on analog charge-sensitive preamplifiers (CSPs) \cite{Abbene:2023vhm, Abbene:2024tpm}. A dedicated digital acquisition system recorded, in real time, the output waveforms from all eight detectors. The acquired data consisted of sets of snapshot waveforms (SWs) per detector channel, where each pulse was time-stamped using the single delay line (SDL) shaping technique.

\noindent The SWs were later analyzed offline within a dedicated LabView-based framework, developed to extract pulse characteristics and generate energy spectra. Each snapshot covered a time window of 10 $\mu$s, which can be considered the actual dead time of the process. This configuration ensures a high throughput, defined as the ratio of output to input counting rate (OCR/ICR), reaching 99\% up to an ICR of 1500 cps, which is the highest rate measured during beam-on conditions. A detailed description of the data acquisition and processing is reported in previous works \cite{Abbene:2023vhm, Abbene:2024tpm}.

\subsection{The DAQ Logic}

\noindent The DAQ logic was based on a setup already employed and described in a previous study \cite{Scordo:2023per}. The signals from the two LUMI scintillators were processed by an ORTEC 935 Constant Fraction Discriminator (CFD) and then sent to a CAEN N93B Mean Timer unit. The output of the Mean Timer was used as the stop signal for an ORTEC 566 Time-to-Amplitude Converter (TAC), while the start signal was provided by a triple coincidence between the DA$\Phi$NE RF/4 clock and the LUMI signals, generated using a CAEN N455 Coincidence Unit. The data in this run were acquired in trigger mode, meaning that the signals on the detectors were recorded each time there was a coincidence between the two TAC and the DA$\Phi$NE collider radiofrequency (RF) signal.

\noindent The TAC output was a signal with an amplitude proportional to the time difference between start and stop, and it was sent to the same digitizer that recorded the CZT signals. Both the TAC and CZT outputs were acquired by a digitizer controlled by a custom firmware developed at the University of Palermo. In addition to waveform acquisition, the digitizer also provided the time of arrival of each signal relative to the TAC output.

\begin{figure} [H]
    \centering
    \includegraphics[width=0.75\linewidth]{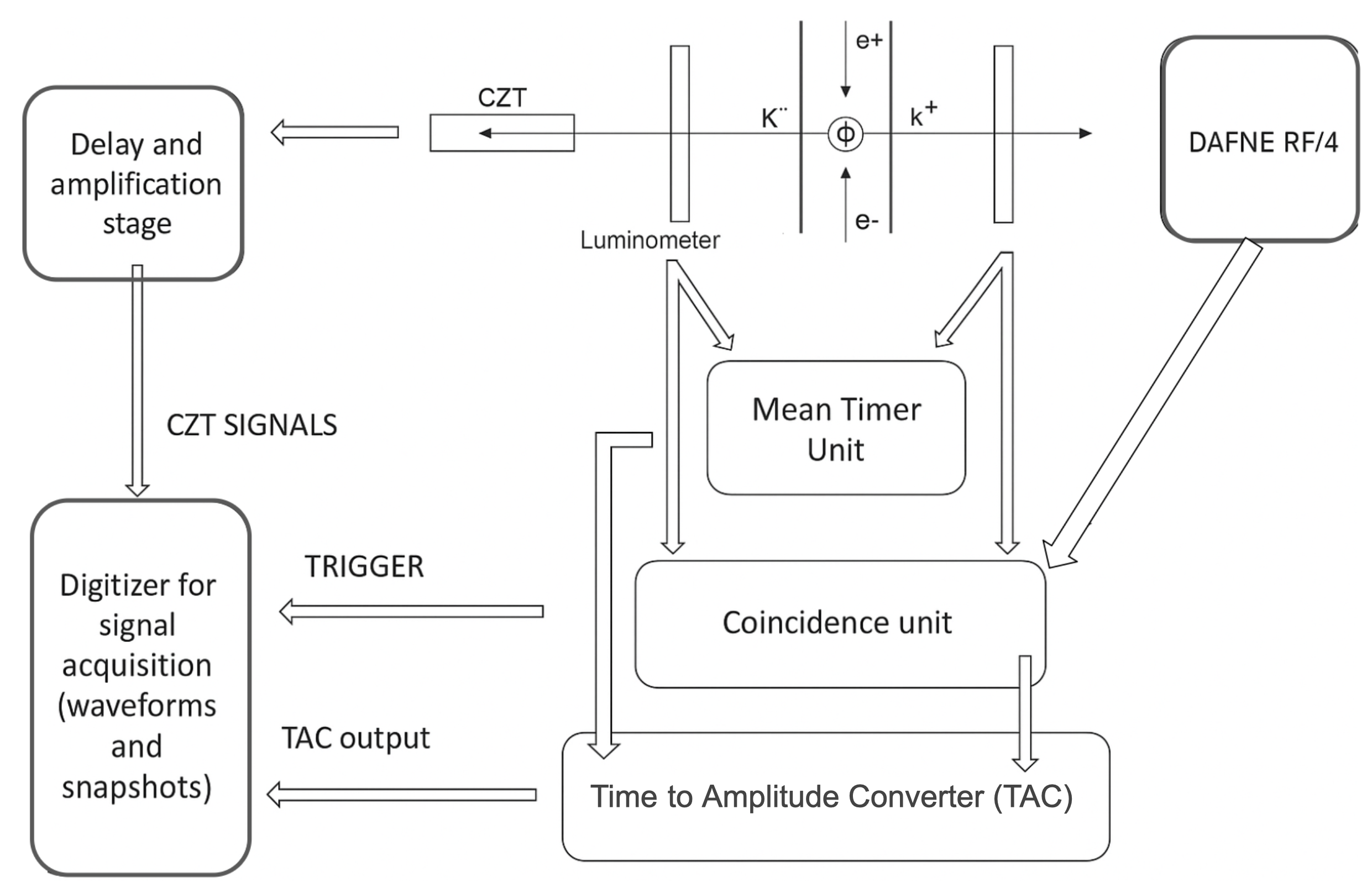}
    \caption{Scheme of the DAQ system of the experiment. Adapted from \cite{Scordo:2023per}.}
    \label{fig: DAQ scheme}
\end{figure}

\noindent A scheme of the DAQ system can be found in Figure \ref{fig: DAQ scheme}.

\section{Data Selection} \label{sec: data analysis}


\noindent This work introduces a data selection strategy that leverages the timing capabilities of CZT-based detectors. Combined with the two LUMI for time-of-flight selection, these detectors proved to be well suited for kaonic atom spectroscopy and, more broadly, for accelerator-based fundamental physics research where background reduction is essential. To this goal, we describe the optimization of data selection performed on the dataset acquired with one of the eight quasi-hemispherical CZT detectors employed in the experiment.

\begin{figure} [H]
    \centering
    \includegraphics[width=0.75\linewidth]{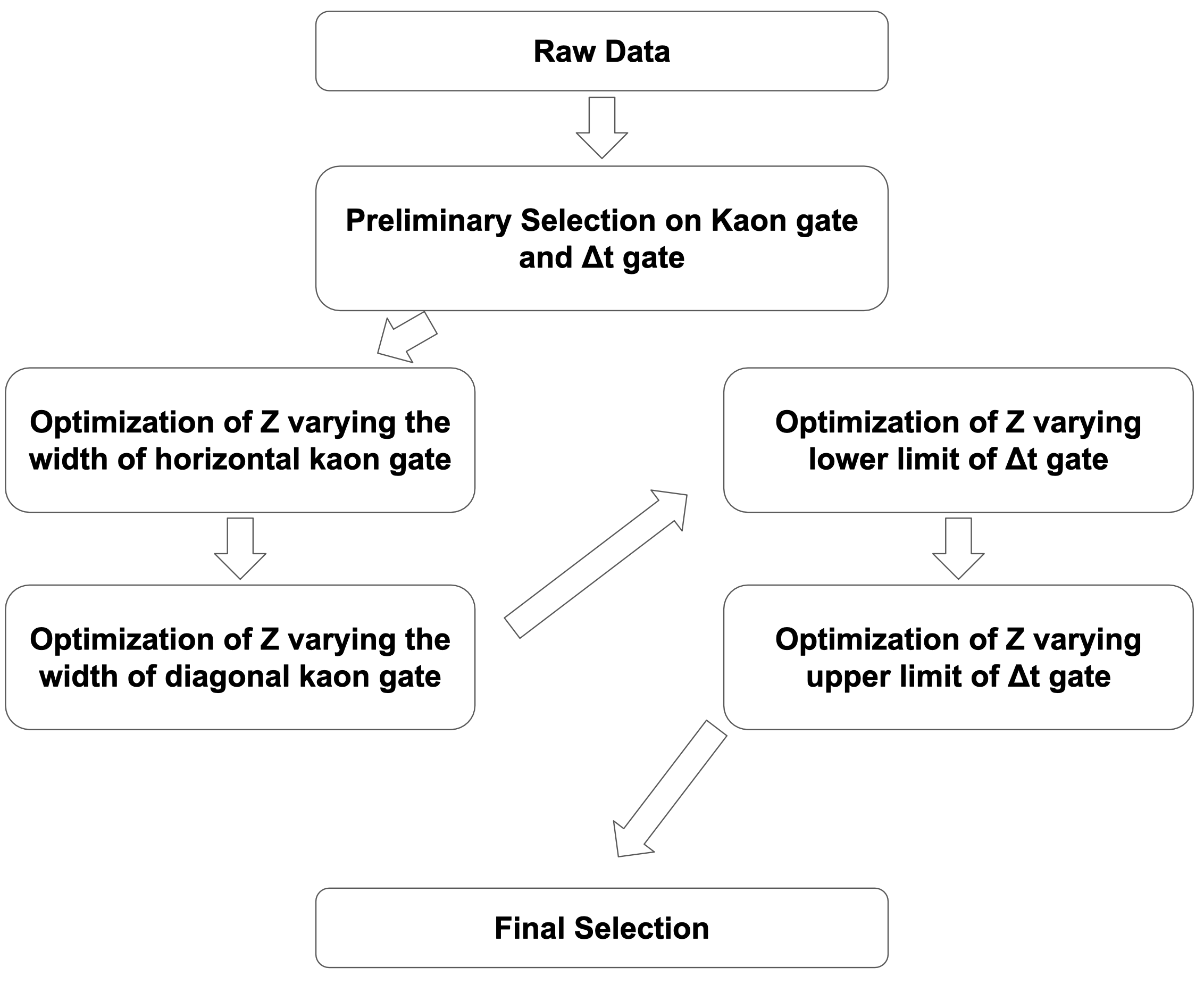}
    \caption{{Scheme of the data selection optimization process described in the section.}}
    \label{fig: data selection flowchart}
\end{figure}

\noindent After summarizing the CZT energy calibration method, we present the approach used to maximize the signal significance (Z) and how it is applied to the selection of kaon-related signals on one of the detectors. This selection is enabled by the LUMI system, as already explored in previous works \cite{Abbene:2023vhm, Scordo:2023per}, and is also used to study and exploit the distribution of the time difference between the detector signal and the trigger ($\Delta t$). Finally, we present the resulting spectrum obtained with the optimized cut and evaluate the background rejection factor achieved through this methodology. { A schematic flowchart of the data selection process can be found in Figure }\ref{fig: data selection flowchart} .

\subsection{Energy Calibration Method}

\noindent {The calibration used for the dataset analysed in this work was performed on March 5$^{\text{th}}$, 2024, in the middle of the data-taking run.} The calibration procedure for the CZT detector has been thoroughly described in a previous work \cite{Abbene:2024tpm}. The detectors were calibrated approximately once every two weeks with the DA$\Phi$NE beam off, using three reference peaks at 39 keV, 121 keV, and 244 keV from a $^{152}$Eu source. 

\noindent The detector calibrated with this method demonstrated good energy stability across the entire data-taking period, confirming the reliability of the system and its resistance to potential issues arising from electronics or environmental conditions.

\noindent For a more detailed discussion of the calibration method, the a priori estimation of systematic errors, and the results on the system’s energy stability, we refer the reader to our previous work \cite{Abbene:2024tpm}.

\subsection{Kaon-MIPs Discrimination} \label{sec: kaon selection}

\noindent The signals from the two LUMI scintillators, combined with the DA$\Phi$NE collider radiofrequency (RF) signal, can be used to discriminate charged kaon pairs originating from the interaction point \cite{Abbene:2023vhm,Scordo:2023per}. The narrow momentum spread and low $\beta$ of the kaons from $\phi$ meson decays allow for the selection of kaon-related signals through time-of-flight measurements. The TAC-processed luminometer used in the experiment was shown to effectively distinguish between kaons and minimum ionizing particles (MIPs) \cite{Abbene:2023vhm,Scordo:2023per}. For the 2024 run, the combined signal from both the boost side and antiboost side LUMIs was employed, representing a significant improvement in kaon pair discrimination capability.

\begin{figure}[H]
    \centering
    \includegraphics[width=0.75\linewidth]{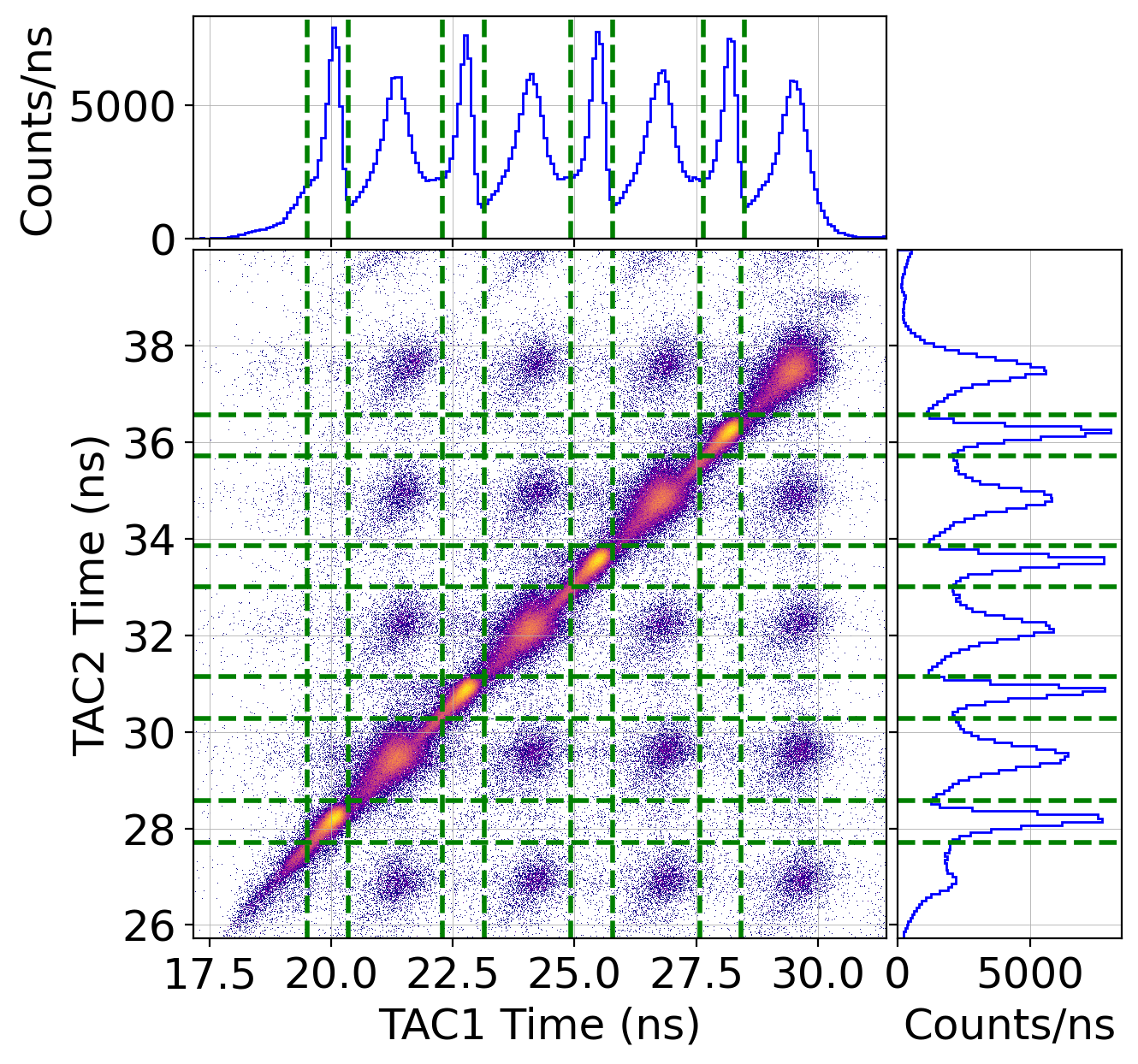}
    \caption{2D histogram of TAC1 vs TAC2. Kaon-related coincidences appear as narrow peaks within the green guidelines.}
    \label{fig: TAC2Drawspectrum}
\end{figure}

\noindent Figure \ref{fig: TAC2Drawspectrum} shows the spectrum of the two luminometer signals after TAC processing and their combined 2D histograms. The narrower peaks correspond to the kaon windows, due to the smaller momentum spread compared to MIPs. Four similar patterns, corresponding to kaon and MIP arrivals, appear because the TAC operates at a frequency of 1/4 of the DA$\Phi$NE RF, which is 3.7 MHz \cite{Milardi:2009zza, Milardi:2018sih, Milardi:2021khj, Milardi:2024efr}.

\noindent In this work, the two time windows for kaon selection were optimized to maximize the signal significance (Z), defined as:
\begin{equation} \label{eq: Z}
\text{Z} = \frac{A_{\text{S}}}{\sqrt{A_{\text{B}}}},
\end{equation}
\noindent
where $A_{\text{S}}$ is the number of events in the Gaussian peak representing the signal, and $A_{\text{B}}$ is the number of background events in the same energy region.

\noindent After a $\Delta t$ cut (see next section), to select only the events on the detector correlated in time with the collisions and remove the asynchronous background, the time-of-flight selection was optimized by calculating the signal significance for different time windows. This yielded a distribution of Z as a function of the selected kaon region, from which the optimal window, corresponding to the maximum Z, was identified. The optimization was performed by fixing the right-hand boundaries of the two TACs (where the signal starts to rise, as shown in Figure~\ref{fig: TAC2Drawspectrum}), while progressively extending the other boundaries to enlarge the kaon window. For each configuration, an energy spectrum was extracted and fitted to evaluate the visible signal, and Z was computed following Equation~\ref{eq: Z}.

\noindent A similar approach was applied using a linear window in the TAC1 vs TAC2 plane, in which the window width was varied while its center remained fixed. This method allowed the identification of the time-of-flight selection region that maximizes signal significance while effectively suppressing background.

\noindent The fitting procedure and details of the signal significance calculation are described below. The number of signal events was determined by fitting the visible kaonic atom transitions in the energy spectrum - specifically the K-Al 5$\rightarrow$4 (50keV) and 4$\rightarrow$3 (106keV) transitions - with Gaussian functions. The means were fixed to values calculated using the MCDFGME code \cite{santos_x-ray_2005}, and the standard deviations were modeled as a function of energy:
\begin{equation}
\sigma(E) = \sqrt{a + bE}
\end{equation}
\noindent following \cite{Gysel:2003}, with the parameters $a$ and $b$ fixed to values determined from detector calibration. The areas of the peaks were left as the only free parameters.

\noindent The background was modeled using a combination of an exponential term, a complementary error function, and a linear term, following the approach in \cite{Abbene:2024tpm}:
\begin{equation} \label{eq: bkg}
f_{\text{bkg}}(x) = a + b \cdot x + c \cdot \exp(d \cdot x) + \text{erfc}\left(\frac{e - x}{f}\right).
\end{equation}
{\noindent where $a$ and $b$ are the parameters of the linear function describing the electronic baseline; $c$ and $d$ are the parameters of the exponential component accounting for the background generated by photons and electrons undergoing Compton scattering; $e$ and $f$ are the two parameters of the complementary error function, which accounts for the low-energy cutoff resulting from the presence of shielding and electronic components. The exponential term for Compton scattering arises from the fact that the particles from the IP interact with the air and the materials between the beam pipe and the detector, producing a considerable number of secondary particles at lower energies.
}
\noindent Two additional background peaks originating from lead fluorescence were included. In particular, the lead K$\alpha$ transition required an extra component to account for incomplete charge collection. This was modelled using an exponential function multiplied by a complementary error function:
\begin{equation}
T(x) = \epsilon \times N \times \exp\left(\frac{x - \mu}{\beta \sigma}\right) \times \text{erfc}\left(\frac{x - \mu}{\sqrt{2}\sigma} + \frac{1}{\sqrt{2}\beta} \right),
\end{equation}
\noindent where $\mu$ and $\sigma$ are the mean and standard deviation of the corresponding Gaussian, $\epsilon$ controls the tail amplitude, and $\beta$ defines its width. This model was introduced and validated in \cite{Abbene:2024tpm}. Tail parameters were fixed to calibration values, and the incomplete charge collection effect was negligible in all other peaks.

\noindent The full background model, including tail components, was also validated during a dedicated run with the DA$\Phi$NE collider operating, as reported in \cite{Abbene:2024tpm}.

\begin{figure} [H]
    \centering
    \includegraphics[width=\linewidth]{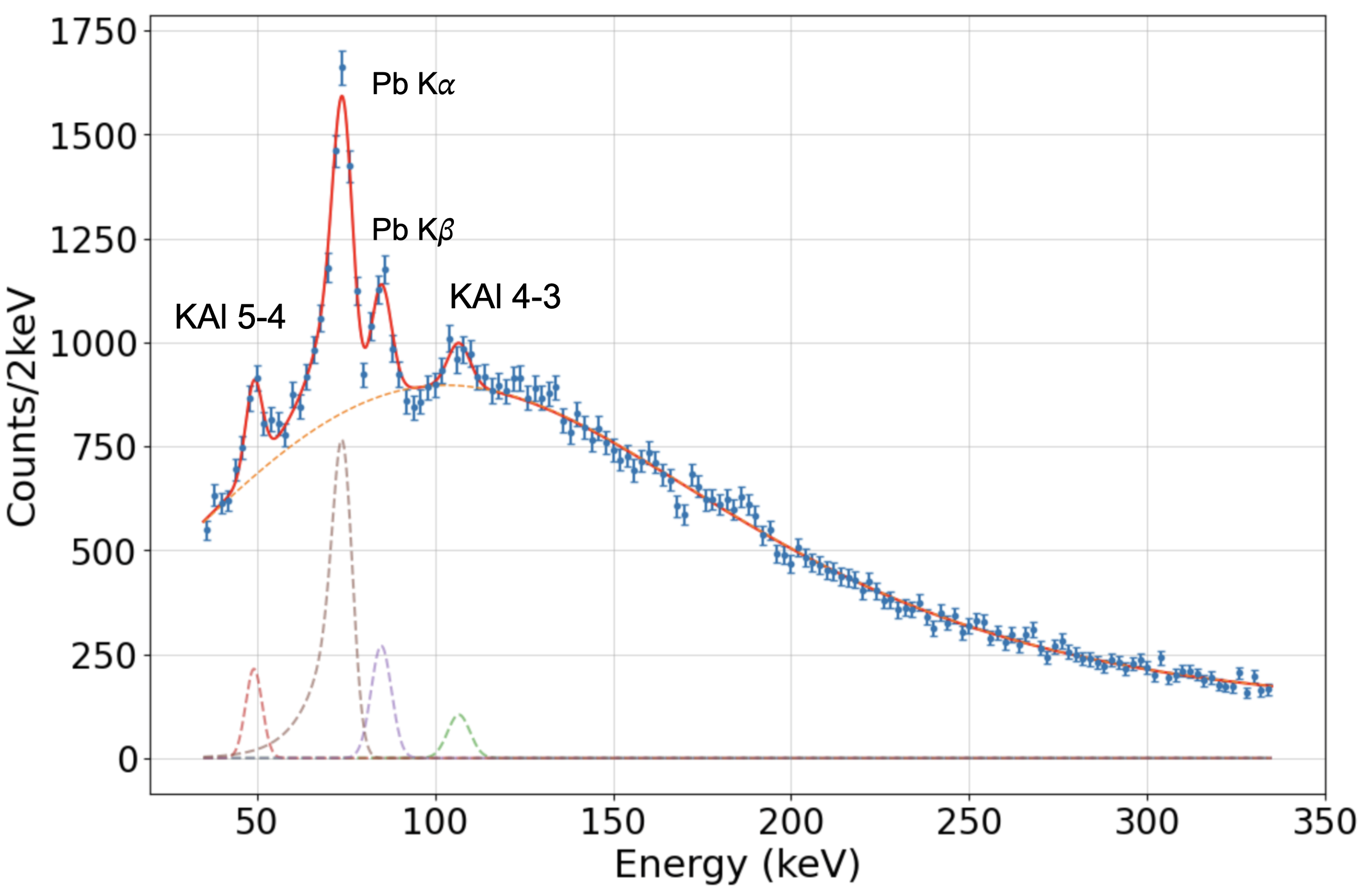}
    \caption{One of the fits to the distribution of events per energy in one CZT detector, used to quantify the signal significance of the kaonic aluminum peaks with the different cuts. The K-Al transitions, the background in Equation \ref{eq: bkg}, and the two background peaks due to lead fluorescence are shown together with the total fit.}
    \label{fig: scan fit}
\end{figure}

\noindent In Figure \ref{fig: scan fit}, a plot showing one of the fits done for optimizing the kaon time-of-flight window is reported, with the KAl and lead transition labelled.

\noindent The Zs for the two K–Al peaks in each time window were then obtained by extracting from the fit the area of the Gaussians representing the signal and the corresponding background area in the region between $\mu - 5 \sigma$ and $\mu + 5 \sigma$ underneath them - with $\mu$ representing the mean of the Gaussian, and $\sigma$ the standard deviation - and finally applying Equation \ref{eq: Z}.

\begin{figure}[H]
    \centering
    \begin{subfigure}[b]{0.48\linewidth}
        \centering
        \includegraphics[width=\linewidth]{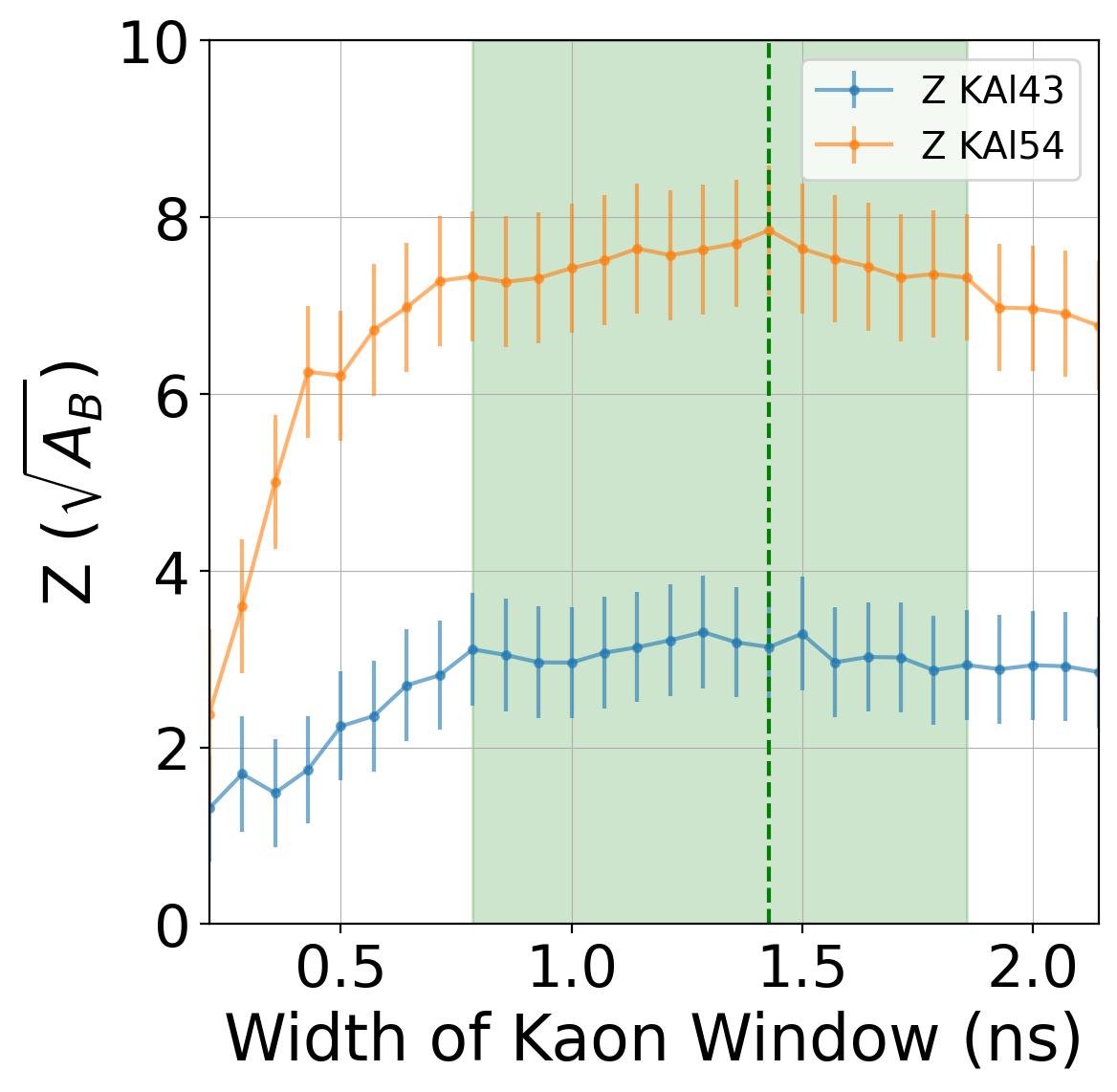}
        \label{fig: horiz tac distr}
    \end{subfigure}
    \hfill
    \begin{subfigure}[b]{0.48\linewidth}
        \centering
        \includegraphics[width=\linewidth]{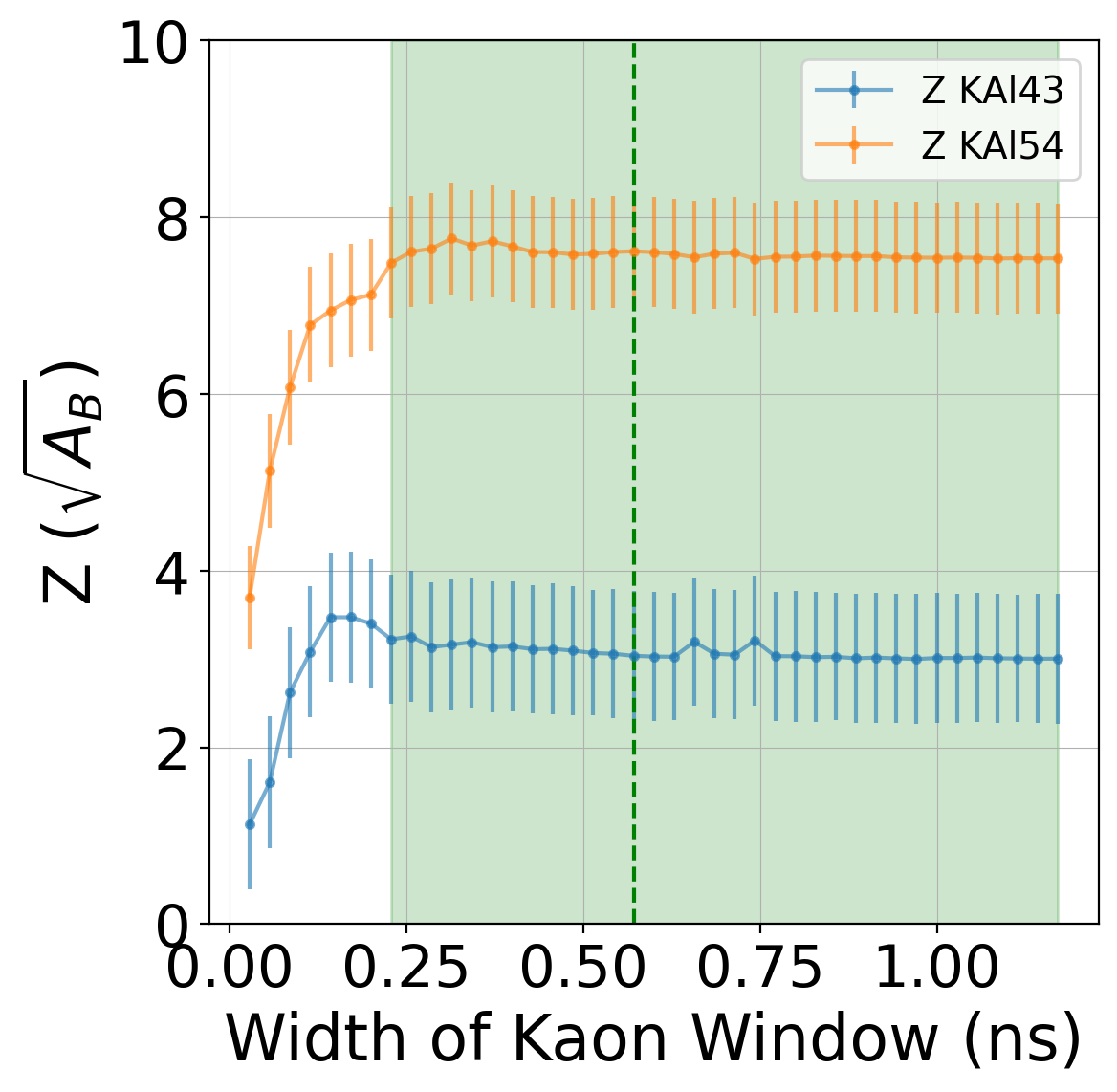}
        \label{fig: diag tac distr}
    \end{subfigure}
    \caption{Left: Z distribution as a function of the TAC kaon selection width, with the selected maximum value (100 ADC TAC) marked by the green line and the maximum region highlighted in green. Right: Z distribution as a function of the selection width in the TAC1 vs TAC2 space, with the selected maximum value (40) marked by the green line and the maximum region highlighted in green. \\ The reported errors correspond to the statistical uncertainties on the signal amplitude.}
    \label{fig:tac_combined_distr}
\end{figure}

\noindent After obtaining the Z distributions as a function of the varied boundary values (as shown in Figure \ref{fig:tac_combined_distr}), an arbitrary maximum Z value within the region of highest values was selected to define the final limit used for the cut. The reported errors correspond to the statistical uncertainties on the signal amplitude, which represent the dominant contribution.

\begin{figure} [H]
    \centering
    \includegraphics[width=\linewidth]{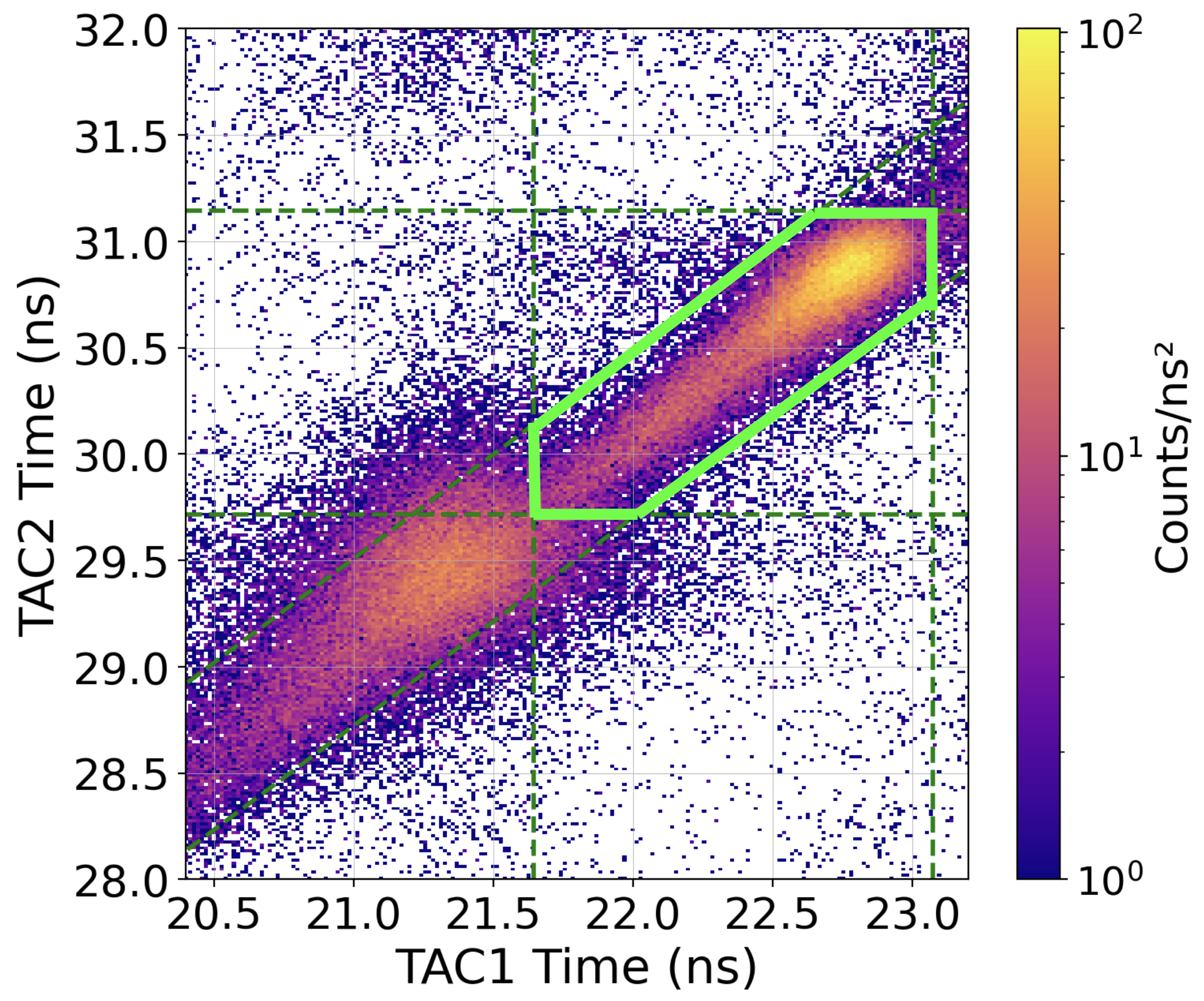}
    \caption{2D histogram of the two TAC signals from the LUMI scintillators for one kaon-MIP couple, with the optimized selection guidelines shown as dotted green lines. The selected kaon windows correspond to the hexagonal areas formed by the broader rectangles along the diagonal, combined with the region inside the diagonal cuts. These selected area is highlighted in light green in the figure.}
    \label{fig: TAC final selection2}
\end{figure}

\noindent The final selection applied to TAC1 and TAC2 for one of the four patterns (see the 2D histogram in Figure \ref{fig: TAC2Drawspectrum}) is shown in Figure~\ref{fig: TAC final selection2}. The green box indicates the optimized region corresponding to the maximum achievable Z, as determined by the procedure described above.


\subsection{Timing Analysis} \label{sec: Dt analysis}

\noindent In kaonic atom spectroscopy at DA$\Phi$NE, as demonstrated by the SIDDHARTA and SIDDHARTA-2 experiments~\cite{SIDDHARTA:2011dsy,Artibani:2024kop,Curceanu:2019uph}, the time distribution of detector signals plays a crucial role in applying time-based cuts. In particular, a clear time correlation exists between the signals originating from kaon production and the collisions, which manifests as a peak in the distribution of the time difference between the trigger and a detector signal ($\Delta t$). By selecting events within this peak region, it is possible to significantly reduce the background from uncorrelated processes, thereby enhancing the signal quality.

\noindent In previous works, CZT detectors demonstrated ideal timing capabilities for kaonic atom measurements. In particular, the characteristic peak in the distribution of the difference between the CZT signal time and the trigger time ($\Delta t$), corresponding to the beam-related events, was observed \cite{Scordo:2023per}. In the run analyzed in this work, the timing distribution proved again to be promising, showing the expected excess in the $\Delta t$ spectrum. 

\begin{figure}[H]
    \centering
    \begin{subfigure}[b]{0.48\linewidth}
        \centering
        \includegraphics[width=\linewidth]{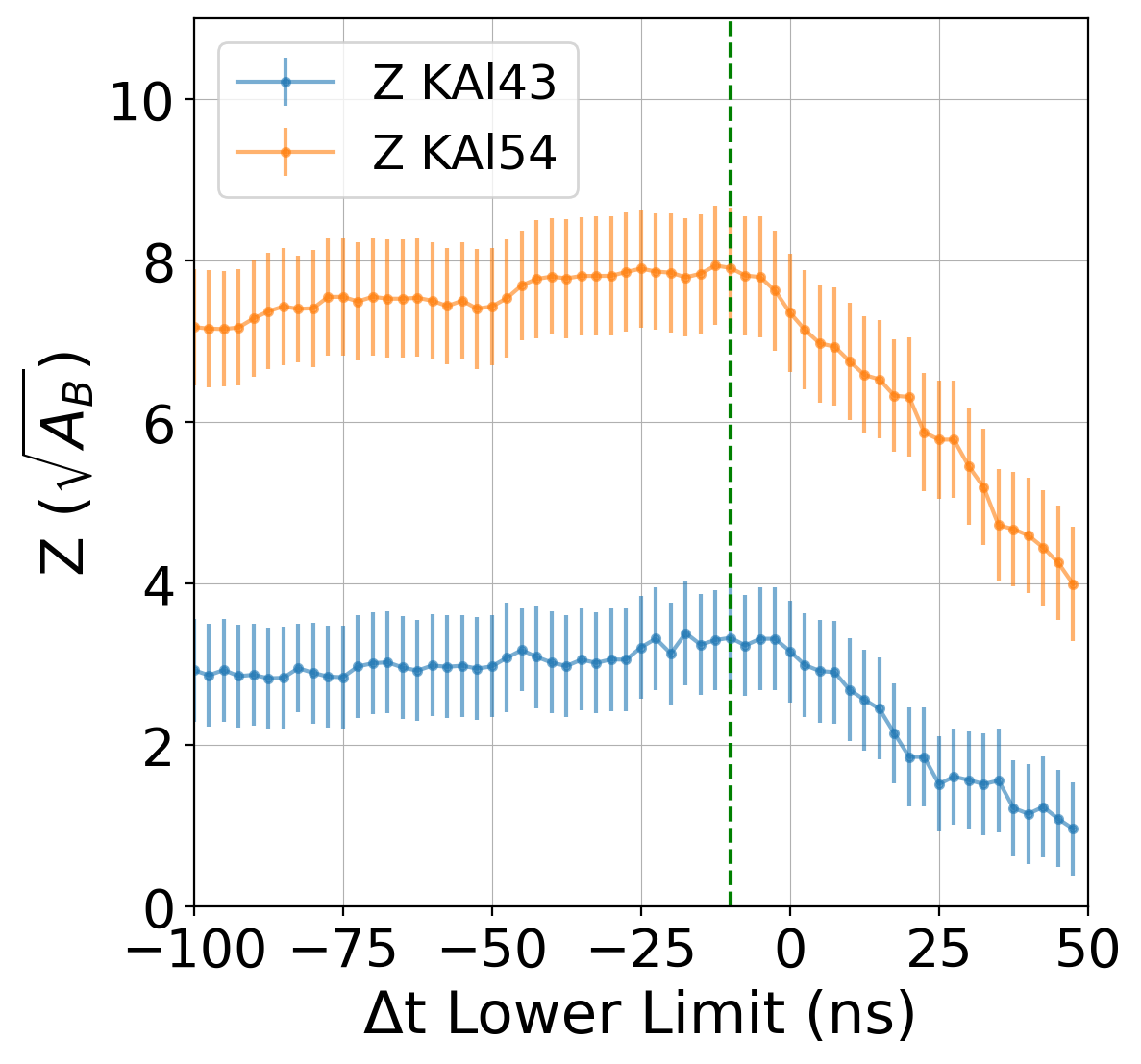}
        \caption{Z vs lower limit of $\Delta t$ (upper limit fixed at -10 ns).}
        \label{fig:scan_dt_low}
    \end{subfigure}
    \hfill
    \begin{subfigure}[b]{0.48\linewidth}
        \centering
        \includegraphics[width=\linewidth]{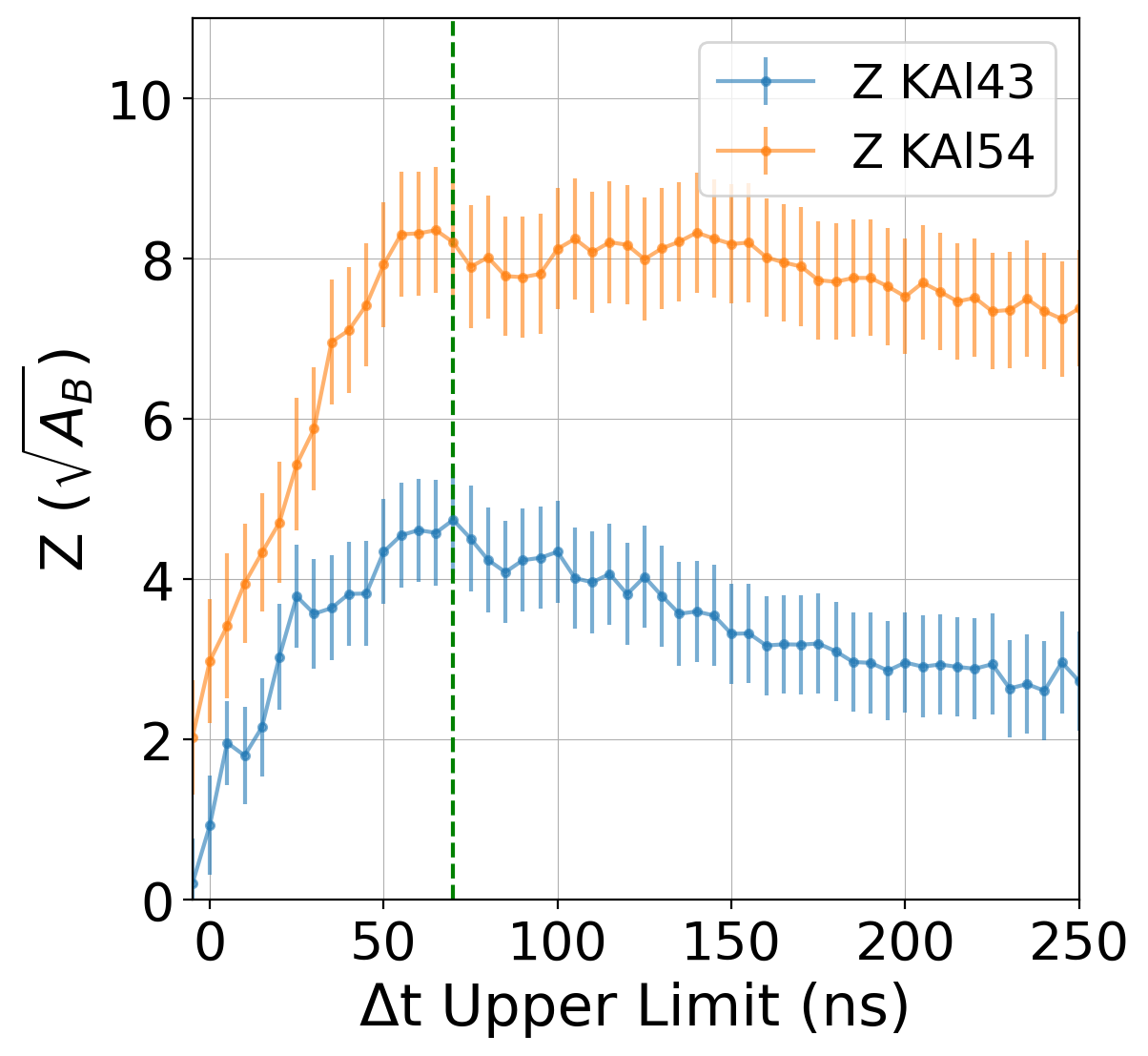}
        \caption{Z vs upper limit of $\Delta t$ (lower limit fixed at 70 ns).}
        \label{fig:scan_dt_up}
    \end{subfigure}
    \caption{Left: Z distribution in function of lower limit of $\Delta t$ with upper limit fixed at 300 ns. The maximum is highlighted in green (-10 ns). Right: Z distribution in function of upper limit of $\Delta t$ with lower limit fixed at 10 ns. The maximum is highlighted in green (70ns).\\ The reported errors correspond to the statistical uncertainties on the signal amplitude.}
    \label{fig:scan_dt_combined}
\end{figure}

\noindent The final $\Delta t$ window was chosen after applying the same method used for the time-of-flight selection, choosing the time window in which the signal significance is higher. At first, the upper limit was fixed to a high value of 300 ns, and the lower limit was varied to obtain the Z distribution. Subsequently, the lower limit was fixed to the optimal value, and the upper limit was varied to obtain the distribution of the Z. The two distributions are reported in Figure \ref{fig:scan_dt_low} and \ref{fig:scan_dt_up}, respectively.

\begin{figure} [H]
    \centering
    \includegraphics[width=\linewidth]{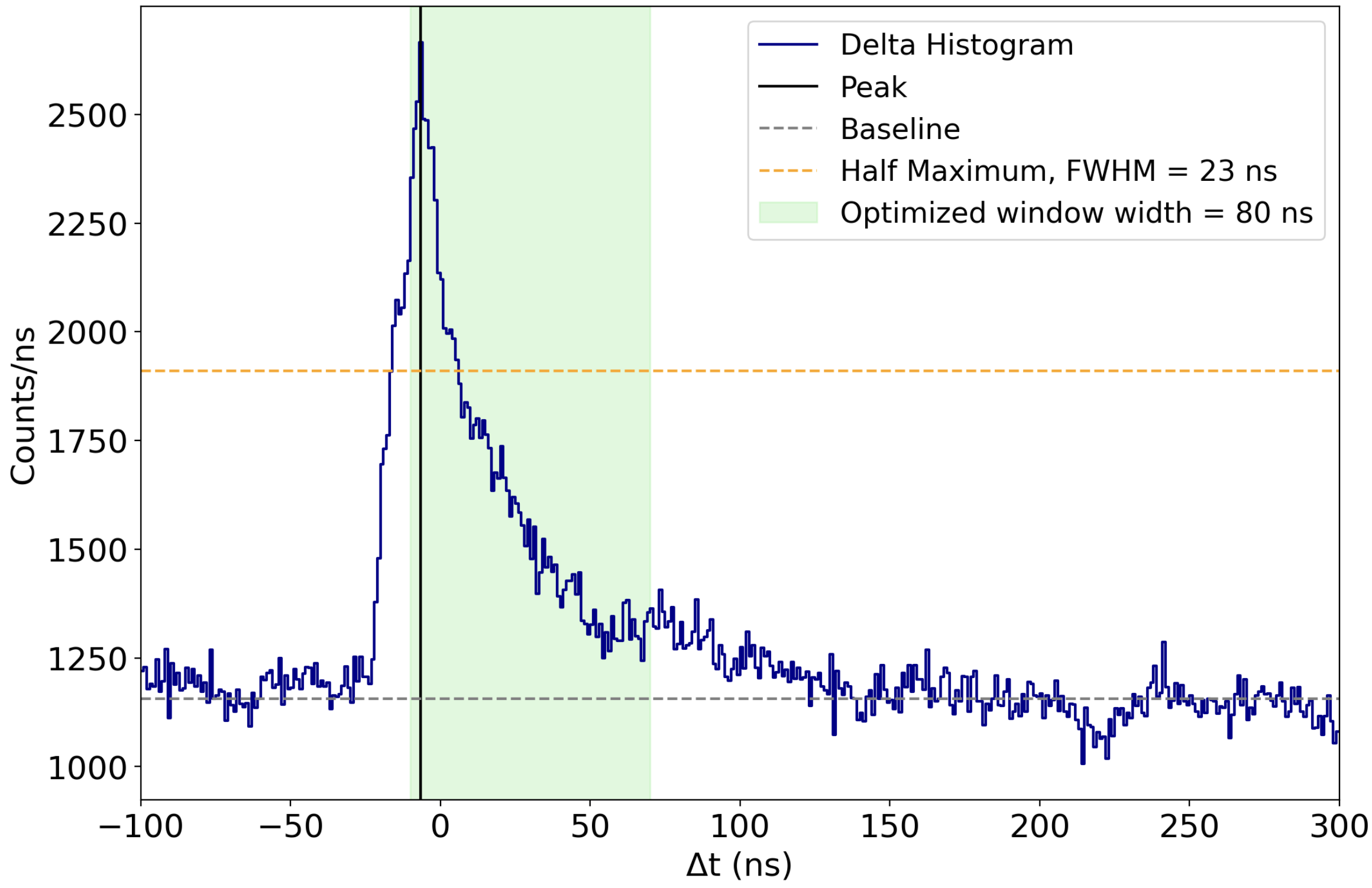}
    \caption{Measured cumulative spectrum of the drift time with the optimized limits of the signal region reported in green. In the legend the FWHM of the peak and the width of the selected window are reported.}
    \label{fig: drift hist with sel}
\end{figure}

\noindent In the first distribution, the maximum value is reached around $\approx$ -10 ns, which corresponds to the chosen lower limit of the window. In the second one, the maximum is found around 70 ns, after which a plateau is observed for the 5-4 transition, less affected by background. This value corresponds to the chosen upper limit of the window. This study confirms that the evident peak around -10 ns is due to the synchronous background, while the signal related to kaonic atoms spans the wider time window. The optimized kaon window limits, highlighted in green, are reported in Figure \ref{fig: drift hist with sel}. 
{The dominant contribution to the observed $\Delta t$ distribution arises from the charge drift time within the CZT detector; a kaon with $\beta \approx 0.25$ covers the 10.2 cm between the interaction point and the target in about 2 ns, while the timescales for kaon capture, atomic cascade, and X-ray emission range from $10^{-12}$ s to $10^{-9}$ s. Likewise, the photon flight time over the 9 cm from the target to the detector is less than 1 ns. The contribution of these processes to the overall $\Delta t$ width is therefore well below the nanosecond level and can be considered negligible. Consequently, the measured FWHM can be regarded as a direct measurement of the detector’s intrinsic time resolution, obtained using X-rays and particles spanning a wide range of energies. A dedicated study performed in a low-background environment with a monochromatic source would provide a more precise characterization.}

\noindent A $\approx$ 100~ns $\Delta t$ window gives a very good result, compared to that achieved with the SDDs used in the experiment {[(507.60 $\pm$ 0.47) ns at FWHM]}  \cite{Sgaramella:2024ehl}. This finding demonstrates the remarkable potential of CZT detector timing for background reduction in high-radiation environments.

\section{Final Selected Spectrum}
\begin{figure} [H]
    \centering
    \includegraphics[width=\linewidth]{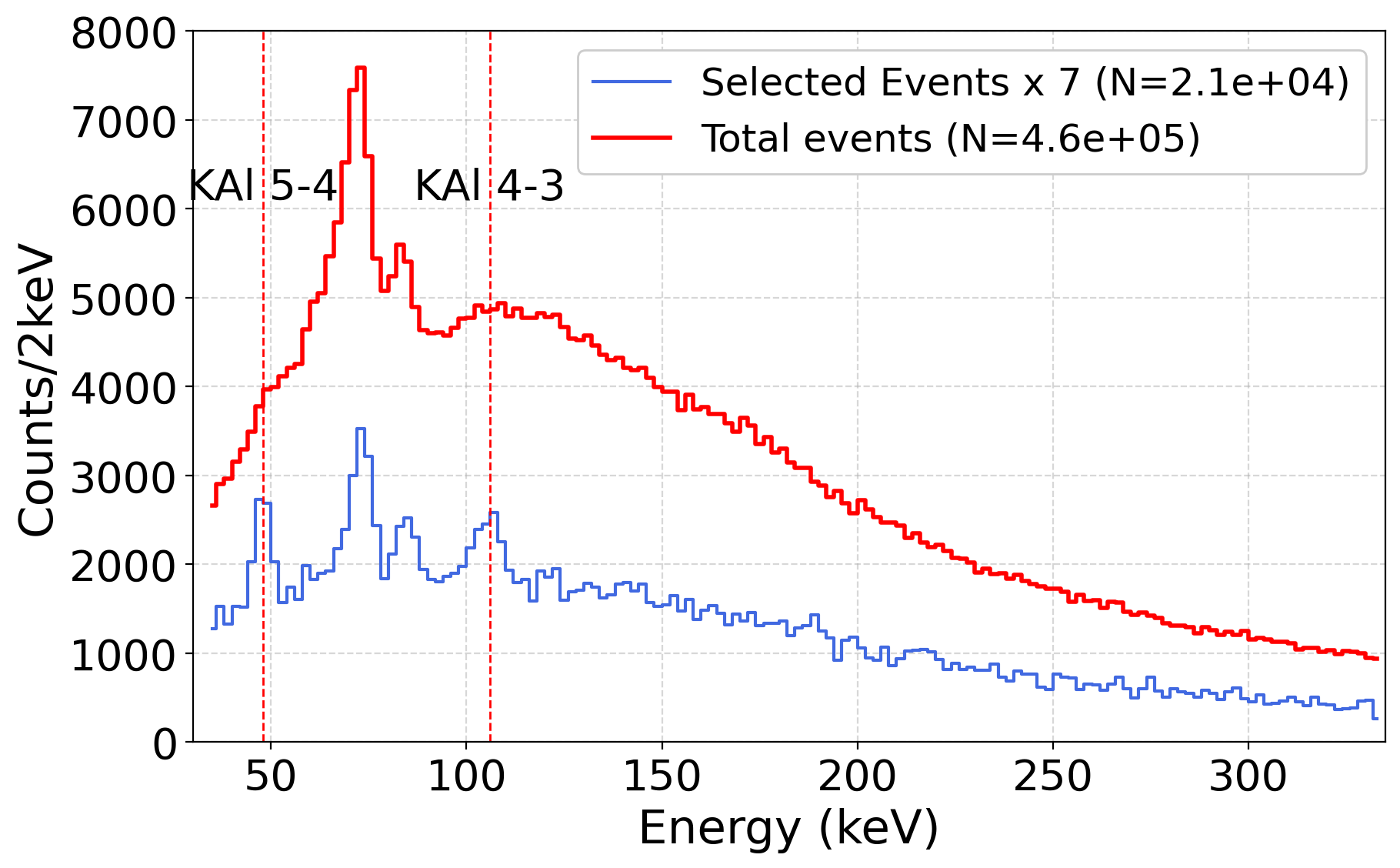}
    \caption{Histogram of the raw spectrum (red), compared to the one after the optimized selections with counts multiplied by a factor 7 (blue), with the kaonic aluminum transitions highlighted in red.}
    \label{fig: final spectrum comp}
\end{figure}

\noindent The spectrum with kaon and $\Delta t$ selections, acquired by the CZT detection system, compared with the raw spectrum, is reported in Figure \ref{fig: final spectrum comp}.

\noindent After the two selection steps, approximately 95\% of the triggered events are discarded. In a previous study~\cite{Abbene:2023ogi}, the overall rejection factor for total events was evaluated using the coincidence of a kaon signal on a single LUMI scintillator and a variable $\Delta t$ window. That work showed that, within a 100 ns $\Delta t$ window, the rejection factor relative to the total number of triggered and untriggered events was on the order of $10^6$, a comparable rejection efficiency is expected when considering the total number of events on the detector. These findings further confirm the method's strong background suppression capability, highlighting its potential for future precision measurements in high-radiation environments.

\noindent In the spectrum, two kaonic aluminum transitions, KAl 5-4 (50 keV), and KAl 4-3 (106 keV), are visible, confirming the robustness of both the analysis and the detection system.

\begin{figure} [H]
    \centering
    \includegraphics[width=\linewidth]{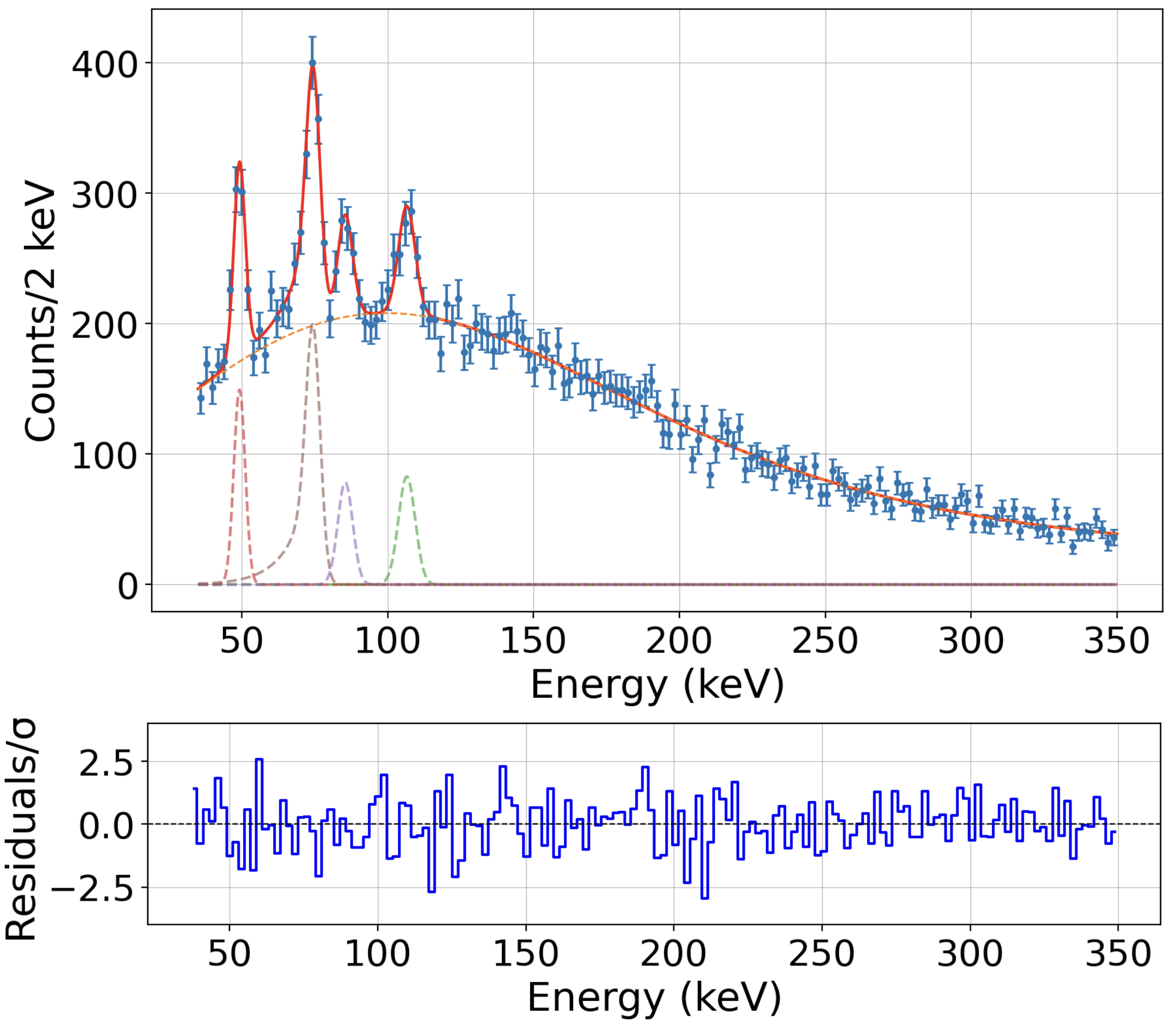}
    \caption{Fit to the data after the optimized selection (up). The errors of the number of counts reported are the statistical errors (1/$\sqrt{counts}$). The plot of residuals for each bin normalized at their standard deviation ($\sigma$) is also reported (down).}
    \label{fig: final fit plot}
\end{figure}

\noindent {\noindent To confirm the successful observation of the two transitions, a final fit was performed on the dataset. The fitting procedure was the same as described in Section~3. In this final fit, the two parameters describing the detector's energy resolution and the two parameters modelling the incomplete charge collection for the Pb~K$\alpha$ transition were left free, in order to avoid bias and to determine the final resolution after the cuts. The resulting fit, together with the residuals, is reported in Figure} \ref{fig: final fit plot}{.}

\noindent {The measured resolutions at 50 keV and 106 keV are respectively (FWHM/nominal value) 9.2 \% 50 keV and 6.3\% at 106 keV, compatible with the ones reported in previous works} \cite{Scordo:2023per, Abbene:2024tpm}.{The resulting peaks and background area with the corresponding Z value are reported in Table} \ref{tab: results events}.


\begin{table}[H]
    \resizebox{\textwidth}{!}{
    \begin{tabular}{c|ccc}
         Transition & \# of Events & \# of Background events in $\pm5 \, \sigma$ & Calculated Z value \\
         ine
        KAl 5-4 & 362 $\pm$ 41 (stat.) $\pm$ 20 (sys.) & 1698 $\pm$ 197 (stat.) $\pm$ 25 (sys.) & 8.78 $\pm$ 1.13 (stat.) $\pm$ 0.49 (sys.) \\
        KAl 4-3 & 295 $\pm$ 50 (stat.) $\pm$ 20 (sys.) & 2939 $\pm$ 500 (stat.) $\pm$ 16 (sys.) & 5.44 $\pm$ 1.03 (stat.) $\pm$ 0.41 (sys.) \\
    \end{tabular}
    }
    \caption{Table reporting the number of signal events, background events and Z value obtained from the final fit fo the two transitions.}
    \label{tab: results events}
\end{table}

\noindent {The systematic uncertainties were evaluated as follows:}
\begin{enumerate}
    \item {Four additional fits were performed by shifting the calibration gain and offset by 0.0015 and 0.075, respectively. According to calibration studies reported in }\cite{Abbene:2024tpm} {, these values correspond to the expected variations over several tens of days. The systematic uncertainty from this effect was estimated by taking the quadratic sum of the deviations from the nominal values after applying the gain and offset shifts.}

    \item {Two additional fits were carried out by varying the energy window limits by $\pm5$~keV (upper limit) and $\pm1$~keV (lower limit), and the resulting differences were used to estimate the related systematic contribution.}

    \item {One additional fit was performed modifying the background model by removing the linear term (parameter $b$ in Equation}~\ref{eq: bkg} {).}

    \item {The binning was varied by $\pm5$ channels, and the corresponding variation in the fitted results was recorded.}

    \item {Finally, the total systematic uncertainty was obtained by adding in quadrature the contributions from all the sources listed above.}
\end{enumerate}

\noindent The observed Z and energy resolution are sufficient to resolve the individual lines, demonstrating the good timing and spectral performance of the CZT detectors, even within the challenging conditions of an accelerator-based experiment.

\noindent This method will be applied to the analysis of the full kaonic aluminum dataset, as well as to the measurements acquired with different targets during the 2024 data-taking campaign, which are expected to provide new results in intermediate-mass kaonic atom spectroscopy.

\section{{Conclusions}}

{\noindent In this paper, we presented the data selection procedure and the first X-ray spectrum acquired by a single CZT detector (out of an array of eight) tested by the SIDDHARTA-2 collaboration during a 20-day run with an aluminum target.

\noindent After briefly describing the experimental setup and calibration methods, we discussed the optimization of the kaon selection window and the time difference ($\Delta t$) window between the trigger and the detector signal, aimed at maximizing the signal significance $Z$.

\noindent We demonstrated that most of the kaonic-aluminum signal is concentrated within a time window of approximately 100~ns from the trigger, confirming the excellent timing performance of CZT detectors for background suppression.

\noindent The resulting energy spectrum, after applying all selection criteria, clearly shows the first observed kaonic-aluminum transition peaks using a room-temperature detector. The significance values calculated using Eq.}~\ref{eq: Z}{ were 5.45~$\pm$~1.03~(stat.)~$\pm$~0.41~(sys.) for the 4–3 transition at 106~keV, and 8.79~$\pm$~1.13~(stat.)~$\pm$~0.49~(sys.) for the 5–4 transition at 50~keV, with corresponding energy resolutions (FWHM/nominal energy) of 9.2\% and 6.3\%, respectively. These results demonstrate not only the effectiveness of the CZT-based detection system developed by the SIDDHARTA-2 collaboration, but also the feasibility of performing precision spectroscopy of kaonic atoms with CZT detectors.

\noindent This work establishes a validated data selection strategy for CZT detectors and represents a proof of concept for future measurements of intermediate-mass kaonic atoms using CZT-based systems. The present analysis was performed on the data acquired by a single detector in a short test run, while the same methodology will be applied to the full dataset collected at DA$\Phi$NE during the 2024 campaign.

\noindent Finally, this study paves the way for new applications of CZT detectors in kaonic-atom spectroscopy, both at DA$\Phi$NE and at J-PARC, demonstrating their strong potential for precision measurements in challenging experimental environments.}

%
%

\section*{Acknowledgements}
{\noindent We thank H. Schneider, L. Stohwasser, and D. Pristauz-Telsnigg from Stefan Meyer-Institut for their fundamental contribution in designing and building the SIDDHARTA-2 setup. We thank as well the INFN, INFN-LNF and the DA$\Phi$NE staff in particular to Dr. Catia Milardi for the excellent working conditions and permanent support. Catalina Curceanu acknowledge University of Adelaide, where part of this work was done (under the George Southgate fellowship, 2024).

Part of this work was supported by the Austrian Science Fund (FWF): [P24756-N20 and P33037-N]; the Croatian Science Foundation under the project HRZZ-IP-2022-10-3878; the EU STRONG-2020 project (Grant Agreement No. 824093); the EU Horizon 2020 project under the MSCA (Grant Agreement 754496); the Japan Society for the Promotion of Science JSPS KAKENHI Grant No. JP18H05402; the SciMat and qLife Priority Research Areas budget under the program Excellence Initiative - Research University at the Jagiellonian University, and the Polish National Agency for Academic Exchange (Grant No. PPN/BIT/2021/1/00037); the EU Horizon 2020 research and innovation programme under project OPSVIO (Grant Agreement No. 101038099). This work was also supported by the Italian Ministry for University and Research (MUR), under PRIN 2022 PNRR project CUP: B53D23024100001.}

\section*{Roles}
{Conceptualization, Alessandro Scordo, Carlo Guaraldo; Methodology, Francesco Artibani and Alessandro Scordo; Software, Francesco Artibani Leonardo Abbene, Antonino Buttacavoli, Fabio Principato and Alessandro Scordo; Formal analysis, Francesco Artibani; Resources, Leonardo Abbene, Manuele Bettelli, Antonino Buttacavoli, Fabio Principato, Andrea Zappettini, Johann Zmeskal and Catalina Curceanu; Data curation, Francesco Artibani, Leonardo Abbene, Antonino Buttacavoli and Alessandro Scordo; Writing – original draft, Francesco Artibani; Writing – review \& editing, Leonardo Abbene, Manuele Bettelli, Antonino Buttacavoli, Fabio Principato, Andrea Zappettini, Massimiliano Bazzi, Giacomo Borghi, Damir Bosnar, Mario Bragadireanu, Marco Carminati, Alberto Clozza, Francesco Clozza, Luca De Paolis, Raffaele Del Grande, Carlo Fiorini, Ivica Friscic, Mihail Iliescu, Masahiko Iwasaki, Aleksander Khreptak, Simone Manti, Johann Marton, Paweł Moskal, Fabrizio Napolitano, Hiroaky Ohnishi, Kristian Piscicchia, Francesco Sgaramella, Michał Silarski, Magdalena Skurzok, Diana Laura Sirghi, Florin Sirghi, Antonio Spallone, Kairo Toho, Lorenzo G. Toscano, Oton Vazquez Doce, Catalina Curceanu and Alessandro Scordo; Supervision, Catalina Curceanu and Alessandro Scordo; Project administration, Catalina Curceanu and Alessandro Scordo.}


\bibliographystyle{elsarticle-num} 
\bibliography{ref.bib}

\end{document}